\documentstyle[12pt,a4]{article}

\def\Journal#1#2#3#4{{#1} {\bf #2}, #3 (#4)}

\def\NCB{\em Nuovo Cimento} 
\def\NPB{{\em Nucl. Phys.} B}

\def\PRD{{\em Phys. Rev.} D}

\def\CQG{\em Class. Quantum Gravity}
\def\IJMPA{{\em Int. J. Mod. Phys.} A}
\def\PJP{\em Pram\=ana Jour. Phys.}

\def\be{\begin{equation}}
\def\ee{\end{equation}}
\def\bea{\begin{eqnarray}}
\def\eea{\end{eqnarray}}

\def\HB{{\cal H}_{\scriptscriptstyle B}}
\def\HQ{{\cal H}_{\scriptscriptstyle Q}}
\def\pb{\phi_{\scriptscriptstyle B}}
\def\pq{\phi_{\scriptscriptstyle Q}}
\def\sz{S_0}
\def\pz{\Phi_0}
\def\tw{T_{\scriptscriptstyle W}}
\def\te{T_{\scriptscriptstyle E}}
\def\tp{T_{\scriptscriptstyle P}}
\def\ts{T_{\scriptscriptstyle S}}
\def\da{\partial a}
\def\db{\partial \pb}
\def\ev{e_v}
\def\rev{\sqrt{\ev a^{2n} + \kappa^2}}
\def\revz{\sqrt{\ev a^{2n} + \kappa_0^2}}
\def\poo{p_{oo}}

\def\Psiz{\Psi_0}
\def\PsiWz{\Psi^{WKB}_0}

\def\sza{{\partial \sz \over \da}}
\def\szb{{\partial \sz \over \db}}
\def\szaa{{\partial^2 \sz \over \da^2}}
\def\szbb{{\partial^2 \sz \over \db^2}}
\def\pha{{\partial \Phi \over \da}}
\def\phb{{\partial \Phi \over \db}}
\def\phza{{\partial \pz \over \da}}
\def\phzb{{\partial \pz \over \db}}
\def\psa{{\partial \psi \over \da}}
\def\psb{{\partial \psi \over \db}}
\def\twa{{\partial \tw \over \da}}
\def\twb{{\partial \tw \over \db}}

\def\tpa{{\partial \tp \over \da}}
\def\tpb{{\partial \tp \over \db}}
\def\dtp{\frac{\partial}{\partial \tp}}

\def\szkz{\sz(\kappa_0)}
\def\pzkz{\pz(\kappa_0)}

\def\szpr{\sz^\prime}

\def\szdpr{\sz^{\prime\prime}}

\def\fpr{f^\prime}
\def\fdpr{f^{\prime\prime}}

\def\sigop{\Sigma_1^+}
\def\sigom{\Sigma_1^-}
\def\sigtp{\Sigma_2^+}
\def\sigtm{\Sigma_2^-}

\def\lp{l_{\scriptscriptstyle P}}
\def\lh{l_{\scriptscriptstyle H}}

\begin{document}
\bibliographystyle{unsrt}
\hfill MMC-M-9 \\
\hspace*{12.5cm} May \ 1996 \\
\begin{center}
{\LARGE Wave Packet in Quantum Cosmology and \\ \vskip 0.5cm
Definition of Semiclassical Time} \\ \vskip 1cm
Y. OHKUWA \footnote{E-mail address: ohkuwa@macqm.miyazaki-med.ac.jp}
\\Department of Mathematics, Miyazaki Medical College,
Kiyotake,\\ Miyazaki 889-16, Japan \\ \vskip 0.3cm 
T. KITAZOE \footnote{E-mail address: kitazoe@hawk.cs.miyazaki-u.ac.jp}
\\Faculty of Engineering, Miyazaki University, Miyazaki 889-21, Japan
\end{center}
\vskip 0.5cm
\begin{abstract}
We consider a quantum cosmology with a massless background 
scalar field $\pb$ and adopt a wave packet as the wave function. 
This wave packet is a superposition of the WKB form wave functions, 
each of which has a definite momentum of the scalar field $\pb$. 
In this model it is shown that  to trace the formalism of 
the WKB time is seriously difficult 
without introducing a complex value for a time.
We define a semiclassical real time variable $\tp$ from the phase of 
the wave packet and calculate it explicitly. 
We find that, when a quantum matter field $\pq$ is coupled to the 
system, an approximate 
Schr\"odinger equation for $\pq$ holds with respect to $\tp$ in a region 
where the size $a$ of the universe  is large and $|\pb|$ is small.  
\end{abstract}

\newpage
\section{Introduction }
The notion of time is one of the most serious problems in quantum 
cosmology$.^{\scriptscriptstyle [1]}$ 
Though this is still controversial, many attempts have been done recently. 
One of them is to utilize the semiclassical approximation. 
Banks and others assumed that the solution to the Wheeler-DeWitt equation 
has the form of the WKB approximation, namely 
$\Psi^{WKB}=\Phi e^{{i \over \hbar}\sz  }$ 
 , where $\sz$ is the Hamilton's principal function, and they introduced 
a time variable $\tw$ , using $\sz$ $.^{\scriptscriptstyle [2]-[13]}$ 
They showed that , when a quantum matter 
field $\pq$ is coupled to the system, its wave function satisfies the 
Schr\"odinger equation with respect to $\tw$ in the region where the 
semiclassical approximation is well justified.

It has not yet been clarified how the classical universe emerged from 
the quantum universe. 
However, it seems probable to assume that the wave function 
of the universe forms a narrow wave packet in the classical region. 
In this paper we will consider a quantum cosmology with a massless background 
scalar field $\pb$ and adopt a wave packet with respect to $\pb$ 
as the wave function. 
This wave packet is a superposition of the WKB form wave functions, 
each of which has a definite momentum $\kappa$ of the scalar field $\pb$ . 
Thus, it is expected that the packet tends to a classical orbital motion 
in the classical region. 
In this model first we will show that it is seriously difficult 
to trace the formalism of 
the WKB time and to define a time variable for the Schr\"odinger equation 
of $\pq$ without introducing a complex value for a time. 

Several years ago Greensite and Padmanabhan advocated 
 a time variable $\te$ by requiring 
that the Ehrenfest principle holds with respect to 
$\te$ $.^{\scriptscriptstyle [14][15]}$
If this Ehrenfest time $\te$ exists, 
this is proportional to another time variable, 
a phase time, which is derived 
from the phase of an arbitrary solution to the Wheeler-DeWitt equation. 
However, recently Brotz and Kiefer showed that the Ehrenfest time $\te$ 
does not always exist, 
since the constraints on $\te$ is considerably 
severe$.^{\scriptscriptstyle [16]}$ 
On the contrary it is possible to define a phase time which is real, 
as long as the phase of the wave function is not constant.  

We will introduce a real background phase time $\tp$   
and calculate it explicitly when the width $\sigma$ of the wave packet 
is narrow and the size $a$ of the universe is large. 
The phase time $\tp$ will be compared to $\tw$ which is derived by a WKB form 
wave function with a momentum $\kappa_0$ where the wave packet has a peak. 
It will be shown that $\tp$ is a smooth extension of $\tw$ and 
they become identical in the narrow limit of the wave packet. 
It is important to  examine whether an expected dynamical equation that is 
a Schr\"odinger equation for $\pq$ with respect to $\tp$ can be derived. 
We  will find that an approximate Schr\"odinger 
equation for $\pq$ holds with respect to $\tp$  in a region 
where the size $a$ of the universe  is large and $|\pb|$ is small.

In \S 2 we first review the WKB approximation and $\tw$, and we construct 
a wave packet from WKB wave functions. 
Next we show that it is seriously difficult to trace the formalism of 
the WKB time and to define a time variable.

In \S 3 we introduce a background phase time $\tp$ , calculate its explicit 
form and compare it with $\tw$ .

In \S 4 we examine whether an approximate Schr\"odinger equation for $\pq$ 
can be derived.

We summarize in \S 5, and the appendix is devoted for the 
detailed estimation of the approximations used in \S 2 and \S 3.

\section{Wave Packet and Difficulty in Definition of Time}
We consider the following minisuperspace model in (n+1)-dimensional space-time
$.^{\scriptscriptstyle [17]}$ 
 Though $n=3$ in reality, we calculate in the more general case. 
The metric is assumed to be
$
ds^2=-N^2(t) dt^2 + a^2(t)d\Omega_n^2 \ ,
$
where $d\Omega_n^2$ is the flat metric.  
We take a massless background scalar field $\pb (t)$ 
and a quantum matter field $\pq (t)$ . 
The Wheeler-DeWitt equation for a wave function $\Psi (a, \pb ,\pq)$ reads 
\bea
{\cal H}\Psi &= &\Bigl( \HB + \HQ  \Bigr) \Psi = 0 \ , \\
\HB &= &{\hbar^2 \over 2v_n a^{n-2}}\biggl({c_n^2 \over a^{\poo}}
{\partial \over \da}a^{\poo}{\partial \over \da} - 
{1 \over a^2}{\partial^2 \over \partial \pb^2} \biggr) + U(a)   \ , \nonumber\\
c_n &= &\sqrt{{{\scriptstyle 16\pi G}\over 2n(n-1)}} \ , \qquad 
U(a) = v_n{2\Lambda \over 16\pi G}a^n \ . \nonumber
\eea
Here $\HQ (a, \pb, \pq)$ is the Hamiltonian constraint for the quantum matter 
field $\pq ,$  
$\poo$ is a parameter of operator ordering, 
$\Lambda$ is a cosmological constant,  
$v_n$ is the spatial volume,  and
we assume that $v_n$ is some properly fixed finite constant.

In order to look for a WKB solution to Eqs. (1) , write 
\be
\Psi( a , \pb , \pq ) = \Phi( a , \pb , \pq) 
e^{{i \over \hbar}\sz ( a , \pb )} \ . 
\ee
Substituting Eq. (2) to Eqs. (1) and equating powers of $\hbar$ , we obtain 
\bea
\openup 8pt
-{c_n^2 \over 2v_n a^{n-2}}\biggl( \sza \biggr)^2 + 
{1 \over 2v_n a^n} \biggl( \szb  \biggr)^2 + 
U(a) &= 0 \ , \\
i\hbar \biggl[  {c_n^2 \over 2v_n a^{n-2}}\biggl( \szaa \Phi +
2\sza \pha + {p_{oo} \over a} \sza \Phi \biggr) \qquad \qquad& \cr
- {1 \over 2v_n a^n}\biggl(\szbb \Phi + 2\szb \phb \biggr) \biggr] + \HQ \Phi  
&= 0 \ , 
\eea
where we have regarded that $\HQ$ is the order of $\hbar$ and 
ignored the terms in the order of $\hbar^2$ . 
The equation (3) is the Hamilton-Jacobi equation for the gravity coupled with a 
background scalar field, and $\sz$ is the Hamilton's principal function.

Let us write the solution $\Phi$ of Eq. (4) as $\pz$ , when there is not the 
quantum matter field $\pq$ , that is 
\be
c_n^2 \biggl( \szaa \pz +
2\sza \phza + {p_{oo} \over a} \sza \pz \biggr) 
- {1 \over a^2}\biggl(\szbb \pz + 2\szb \phzb \biggr)  = 0  \ .
\ee
Now we write 
\be
\Psi( a , \pb , \pq ) = \pz( a , \pb ) \ e^{{i \over \hbar}\sz ( a , \pb )} \ 
\psi ( a , \pb , \pq )\ .
\ee
Then from Eqs. (4)-(6) we obtain 
\be
i\hbar \biggl[  {c_n^2 \over v_n a^{n-2}}\sza \psa  
- {1 \over v_n a^n}\szb \psb  \biggr] + \HQ \psi = 0 \ .
\ee
If we define a time variable $\tw$ as
\be
\openup 6pt
-{c_n^2 \over v_n a^{n-2}}\sza \twa  
+ {1 \over v_n a^n}\szb \twb = 1 \ ,
\ee
Eq. (7) can be written as 
\be
i\hbar {\partial \psi \over \partial \tw} = \HQ \psi \ .
\ee
This is a Schr\"odinger equation, so $\tw$ is a semiclassical 
time variable in the WKB approximation$.^{\scriptscriptstyle [2]-[13]}$  

As in Ref. [17] 
Eqs. (3),(5),(8) can be solved by the separation of variables, 
and solutions are as follows. 
\bea
\sz &= &{\epsilon_a \over c_n}I_S + \kappa \pb + const. \ , \\
I_S &= &{1 \over n} \biggl[ \rev + {\kappa \over 2} 
{\rm ln} \biggl({\rev - \kappa \over \rev + \kappa}\biggr) \biggr] \ , 
\nonumber\\
\pz &= &c_\phi a^{1-\poo \over 2} (\ev a^{2n} + \kappa^2)^{-{1 \over 4}} \ 
\qquad\qquad\qquad\qquad\qquad (\gamma=0) \ ,\\
&= &a^{1-\poo \over 2} (\ev a^{2n} + \kappa^2)^{-{1 \over 4}} \times 
\nonumber \\
&  &\times \biggl\{ c_\phi -{(n-1+\poo)\gamma \over 4\kappa}\pb 
I_\Phi[{\scriptstyle {1\over 4}}]
+{n\gamma \kappa \over4}\pb I_\Phi[{\scriptstyle {-{3\over 4}}}]
+{\epsilon_a \gamma \over 2c_n}
I_\Phi[{\scriptstyle {-{1\over 4}}}] \biggr\} \ \  \nonumber\\
& &+ \frac{\gamma \pb}{2\kappa} 
\qquad\qquad\qquad\qquad\qquad\qquad\qquad\qquad\qquad (\gamma\neq 0) \ , \\ 
I_\Phi[x] &= &\int da \ a^{\poo-3 \over 2}(\ev a^{2n} + \kappa^2)^x \ , 
\nonumber \\ 
\tw &= &- {\epsilon_a v_n \over c_n \sqrt{\ev}} {\rm ln} \ a \ + \ 
\tau_{\scriptscriptstyle W} (\pb) 
\qquad\qquad\qquad\qquad\qquad (\kappa=0)\ , \\
&= &{\epsilon_a  \over c_n}(\xi I_{W 1}-v_n I_{W 2}) \ +\ 
{\xi \over \kappa}\pb \ + \ const. \qquad\qquad (\kappa \neq 0) \ , \\
I_{W 1} &= &{1 \over 2n\kappa}{\rm ln} 
\biggl({\rev - \kappa \over \rev + \kappa} \biggr) \ , \nonumber\\
I_{W 2} &= &-{1 \over 2n\sqrt{\ev}}{\rm ln} \Biggl( 
{\rev-\sqrt{\ev} a^n \over \rev + \sqrt{\ev} a^n}\biggr) \ , \nonumber\\
\nonumber
\eea
where $ \epsilon_a = \pm 1 , \ev = 4v_n^2 \Lambda / 16\pi G $ , 
$ \tau_W (\pb) $ is any function of $\pb$ and
$ \kappa , \gamma , \xi ,  c_\phi $ are arbitrary constants. 
We can identify $\kappa$ as the momentum of $\pb$ . 

This WKB time $\tw$ is  very natural  as a semiclassical time variable 
in the region where the WKB approximation is well justified 
$.^{\scriptscriptstyle [11]-[13]}$  
However, this formalism crucially depends on the assumption that 
the wave function has the WKB form (2) . 
How can we define a time variable, if a wave function $\Psi$ is a superposition 
of the WKB form wave functions, for example, $\Psi$ is the following 
Gaussian wave packet?

\bea
\Psi &= &\Psiz \ \psi \ , \qquad\qquad\qquad\qquad \quad\quad  \\
\Psiz &= &\int d \kappa A(\kappa) \PsiWz (\kappa)  , \\
A(\kappa) &= &\frac{1}{\sqrt{2\pi}\sigma} 
{\rm exp} \Bigl[ -\frac{(\kappa-\kappa_0)^2}{2\sigma^2} \Bigr] , 
\qquad \PsiWz (\kappa)  = 
\pz (\kappa)  e^{{i \over \hbar}\sz (\kappa)} \ ,  \nonumber \\ 
\nonumber
\eea
where $\kappa_0 $ and $\sigma$ are arbitrary constants, 
$\pz (\kappa) , \sz (\kappa)$ are given in Eqs. (10) - (12), 
and $\psi = \psi (a, \pb, \pq) $ is a wave function 
for the quantum matter field $\pq$ . 
Note that, in the limit $\sigma \rightarrow 0 $ , 
$\Psi_0$ becomes identical to $\Psi^{WKB}_0$ . 

First let us try to trace the formalism of the WKB time and to 
define a time variable for the wave packet. 
Substituting Eqs. (15), (16) to Ees. (1) and using Eqs. (3), (5) , 
we obtain 
\be
\int d \kappa A \pz   e^{{i \over \hbar}\sz}
\biggl[ i\hbar \biggl(  {c_n^2 \over v_n a^{n-2}}\sza \psa  
- {1 \over v_n a^n}\szb \psb  \biggr) + \HQ \psi \biggr] = 0 \ ,
\ee
where we have neglected the  higher order terms in $\hbar$ . 
This gives
\bea
i\hbar {\partial \psi \over \partial \ts} &= &\HQ \psi \ , \\
\frac{\partial}{\partial \ts} 
&= &\underline{\int d \kappa A \PsiWz  
\biggl( - {c_n^2 \over v_n a^{n-2}}\sza \frac{\partial}{\partial a}  
+ {1 \over v_n a^n}\szb \frac{\partial}{\partial \pb}  \biggr) }  \\
&  &\qquad  \int d \kappa A \PsiWz  
\qquad\qquad\qquad\qquad\qquad\qquad\qquad  . \nonumber
\eea
At a glance this seems a Schr\"odinger equation with a time $\ts$ . 
However, $\ts$ can not be a real variable in general. 
Therefore it is difficult to define a time variable in this way.

\section{Phase Time}
Next let us introduce a background phase time $\tp$ . We write 
the background wave packet $\Psiz$ in Eqs. (16) as 
\be
\Psiz = \rho e^{{i \over \hbar} \theta }  \ , 
\ee
where $\rho$ and $\theta$ are real, that is $\rho$ and $\frac{\theta}{\hbar}$ 
are the absolute value and the phase of $\Psiz$ , respectively. 
If we replace $\sz$ in Eq. (8) with $\theta$ 
and define a background phase time $\tp$ as 
\begin{equation}
 -{c_n^2 \over v_n a^{n-2}}{\partial \theta \over \partial a} \tpa  
+ {1 \over v_n a^n}{\partial \theta \over \partial \pb} \tpb = 1 \ ,
\end{equation}
we can obtain a real time variable $\tp$ . 

However, the problem is whether an expected dynamical equation that is 
a Schr\"odinger equation for $\pq$ with respect to $\tp$ can be 
derived or not. Note that our background phase time $\tp$ is a little 
different from the phase time which is proportional to the Ehrenfest time 
$\te$ $.^{\scriptscriptstyle [14]-[16]} $
The latter phase time should be defined by the all over phase of 
a solution $\Psi$ 
to the Wheeler-DeWitt equation and depend on not only background fields 
$a , \pb$ but also the quantum matter field $\pq$ . 
Before we examine the Schr\"odinger equation, let us calculate $\tp$ explicitly 
and compare it with $\tw$ .

For simplicity we choose $ \poo = 1 $ , $\gamma = 0$ . 
Then Eqs. (16) can be written as 
\bea
\Psiz &= &c_{\phi} \int d \kappa A (\kappa) e^{f(\kappa)} \ , \\ 
f(\kappa) &= &\varphi (\kappa) + \frac{i}{\hbar}\sz(\kappa) \ , \qquad
\varphi(\kappa) = -\frac{1}{4} {\rm ln} (\ev a^{2n} + \kappa^2) \ , 
\nonumber
\eea
where the last equation means $ \Phi_0 = c_{\phi} e^{\varphi}$  \ . 
If we assume that $A(\kappa)$ has a narrow peak at $\kappa_0$ \ , 
namely $\sigma$ is small, then $\sz$ and $\varphi$ can be expanded 
around $\kappa_0$ \ . 
We neglect higher terms than $(\kappa-\kappa_0)^2$  
and integrate with respect to $\kappa$ \ , and we obtain
\begin{equation}
\Psiz  =  \frac{c_\phi}{\sqrt{1 - \sigma^2 \fdpr}} 
{\rm exp} 
\biggl[ f + \frac{\sigma^2 f^{\prime 2}}{2(1 - \sigma^2 \fdpr)} \biggr] \ , 
\end{equation}
where $f$ means $f(\kappa_0)$ and prime means a partial derivative with 
respect to $\kappa$ namely  
$\fpr = \frac{\partial f}{\partial \kappa}(\kappa_0)$ .

Suppose we assume $\sigma^2$ is small enough to satisfy
\be
1 \gg \sigma^2 | \fdpr | \  
\ee
and neglect higher terms than $\sigma^2$ \ , we have
\begin{equation}
\Psiz  =  c_\phi 
{\rm exp} \Bigl[ f + \frac{\sigma^2}{2} ( f^{\prime 2} + \fdpr ) \Bigr] \ .
\end{equation}

Let us consider the case when the size $a$ of the universe is large enough 
to satisfy following conditions,
\bea
|S_0^{\prime\prime}  | &\gg &| \varphi^\prime  S_0^\prime | \ ,  \\
S_0^{\prime 2} &\gg &\hbar^2 | \varphi^{\prime 2} 
+ \varphi^{\prime\prime}   | \ .
\eea
Detailed estimation of these conditions is given in the appendix. 
In this case $\rho$ and $\theta$ in Eq. (20) can be written as  
\bea
\rho &= &\pzkz 
{\rm exp} \Bigl[ -\frac{\sigma^2}{2\hbar^2} \szpr (\kappa_0)^2 \Bigr] \ , \\ 
\theta &= &\szkz + \frac{\sigma^2}{2} \szdpr (\kappa_0) \ ,
\eea
where we have ignored higher order infinitesimals. 

At this stage it is interesting to mention a relation between 
the trace of the wave packet peak and the classical path. 
When $\sz$ in Eq. (10) is written as 
$
\sz = {\epsilon_a \over c_n}I_S (a, \kappa) + \kappa \; (\pb-\alpha) 
$
with an arbitrary constant $\alpha$,
the classical path is derived from 
\be
\szpr = \frac{\partial \sz}{\partial \kappa} = \beta 
\ee
according to the Hamilton-Jacobi theory. 
Since the parameter $\beta$ is additional and can be absorbed by $\alpha$ , 
we can set as 
\be
\szpr = 0 
\ee 
in this case. The peak of the wave function is obtainable from Eq. (28) . 
If we take the lowest order WKB approximation, we arrive at the classical 
path equation (31) by setting 
$\szpr (\kappa_0 ) = 0$ as a peak of the amplitude $\rho$ .

>From Eqs. (10),(29) we find Eq. (21) becomes 
\be
-\frac{\epsilon_a c_n}{v_n a^{n-1}}
\frac{[\ev^2 a^{4n} + \ev (2 \kappa_0^2 + \frac{\sigma^2}{2}) a^{2n} 
+ \kappa_0^4 ]}
{( \ev a^{2n} + \kappa_0^2  )^{\frac{3}{2}}  }
\tpa  + {\kappa_0 \over v_n a^n} \tpb = 1 \ ,
\ee
and this can be solved by the separation of variables. 
After some calculation the result of $\tp$ is
\bea
\tp &= &- \frac{\epsilon_a v_n}{2 n c_n \sqrt{\ev}} 
{\rm ln} \Bigl( a^{2n} + \frac{\sigma^2}{2 \ev} \Bigr) 
+ \tau_P (\pb)  \qquad\qquad\  (\kappa_0 = 0) \ ,  \\
&= &\frac{\epsilon_a}{c_n} ( \zeta I_{P1} 
- v_n I_{P2} ) 
+ \frac{\zeta}{\kappa_0} \pb + const. \qquad\qquad\ \  (\kappa_0 \neq 0) \ , \\
I_{P1} &= &\int da  \frac{( \ev a^{2n} + \kappa_0^2  )^{\frac{3}{2}}}
{a [\ev^2 a^{4n} + \ev (2 \kappa_0^2 + \frac{\sigma^2}{2}) a^{2n} 
+ \kappa_0^4 ]} \ ,  \nonumber \\
&= &-\frac{1}{n} \biggl[  
\frac{(\sigop)^\frac{3}{2}}{2(\sigop-\sigom)(\kappa_0^2-\sigop)} 
{\rm ln} \frac{\alpha_1-\sqrt{\scriptstyle\sigop}}
{\alpha_1+\sqrt{\scriptstyle\sigop}}  \nonumber \\
& & + \frac{(-\sigom)^\frac{3}{2}}{(\sigop-\sigom)(\sigom-\kappa_0^2)} 
{\rm arctan} \frac{\alpha_1}{\sqrt{-\scriptstyle \sigom}} 
- \frac{1}{2\kappa_0} {\rm ln} \frac{\alpha_1-\kappa_0}{\alpha_1+\kappa_0}
\biggr]  \ , \nonumber\\
I_{P2} &= &\int da  \frac{ a^{n-1} ( \ev a^{2n} + \kappa_0^2  )^{\frac{3}{2}}}
{ [\ev^2 a^{4n} + \ev (2 \kappa_0^2 + \frac{\sigma^2}{2}) a^{2n} 
+ \kappa_0^4 ]} \ ,  \nonumber \\
&= &\frac{1}{n} \biggl[  
\frac{(\sigtp)^\frac{3}{2}}{2(\sigtp-\sigtm)(\ev-\sigtp)} 
{\rm ln} \frac{\alpha_2-\sqrt{\scriptstyle\sigtp}}
{\alpha_2+\sqrt{\scriptstyle\sigtp}}  \nonumber \\
& & + \frac{(-\sigtm)^\frac{3}{2}}{(\sigtp-\sigtm)(\sigtm-\ev)} 
{\rm arctan} \frac{\alpha_2}{\sqrt{-\scriptstyle \sigtm}} 
- \frac{1}{2\sqrt{\ev}} 
{\rm ln} \frac{\alpha_2-\sqrt{\ev}}{\alpha_2+\sqrt{\ev}}
\biggr]      \ , \nonumber 
\eea
where 
$\tau_P(\pb)$ is an arbitrary function of $\pb , $  
 $\zeta$ is any constant in the separation of variables, 
$\alpha_1 = \sqrt{\ev a^{2n} + \kappa_0^2} , \ 
\alpha_2 = \sqrt{\ev+ \kappa_0^2 a^{-2n}}$ , 
$ \Sigma_1^\pm = (-\sigma^2 \pm 
\sqrt{\sigma^4 +8\sigma^2\kappa_0^2})/4 , $ 
$ \Sigma_2^\pm = (-\sigma^2\ev \pm 
\sqrt{\sigma^4  +8\sigma^2\kappa_0^2}\ \ev)/4\kappa_0^2$ , 
 and we have assumed $ \Lambda > 0 $ . 
We find from Eqs. (13),(14) and (33),(34) that, 
when $\kappa$ in Eqs. (13),(14) is replaced by $\kappa_0$,  
$ \xi=\zeta$ , $\tau_W = \tau_P$ 
and $\sigma\rightarrow 0 $, $\tp$ becomes equal to $\tw$ , 
which may be expected from the fact that $\Psiz$ becomes identical to 
$\PsiWz$ when $\sigma\rightarrow 0 $ .

\section{Schr\"odinger Equation}
We will now examine a Schr\"odinger equation for $\pq$ . 
Suppose we substitute Eq. (15) into Eqs. (1), we obtain
\bea
\frac{\hbar^2}{v_n a^{n-2}} \Biggl[
c_n^2 \Biggl( \frac{1}{\Psiz}\frac{\partial \Psiz}{\da} 
+ \frac{1}{2a}  \Biggr) \psa 
&-&\frac{1}{a^2 \Psiz}\frac{\partial \Psiz}{\db} \psb \nonumber \\
&+ &\frac{1}{2} \Biggl( c_n^2 \frac{\partial^2 \psi}{\partial a^2}
-\frac{1}{a^2} \frac{\partial^2 \psi}{\partial \pb^2}  \Biggr) \Biggr] 
+\HQ \psi = 0 \ , 
\eea
where we have used the WKB approximation for the background wave function, 
that is $\  \HB \Psiz = 0$ . Using Eqs. (20) and (21), we can derive 
\bea
i \hbar\frac{\partial \psi}{\partial \tp}&=&\HQ \psi
+ \Biggl( F_a \frac{\partial}{\da} + F_\phi\frac{\partial}{\db} \Biggr) 
\psi + \frac{\hbar^2}{2v_n a^{n-2}}
\Biggl( c_n^2 \frac{\partial^2 \psi}{\partial a^2}
-\frac{1}{a^2} \frac{\partial^2 \psi}{\partial \pb^2}  \Biggr) \ , \\
\dtp&=&-\frac{c_n^2}{v_n a^{n-2}} 
\frac{\partial \theta}{\da}\frac{\partial}{\da}
+\frac{1}{v_n a^n}
\frac{\partial \theta}{\db}\frac{\partial}{\db} \ , \nonumber \\
F_a&=&\frac{\hbar^2 c_n^2}{v_n a^{n-2}} 
\biggl( \frac{1}{\rho} \frac{\partial \rho}{\da}
+ \frac{1}{2a} \biggr) \ , \nonumber \\
F_\phi&=&-\frac{\hbar^2}{v_n a^n}\frac{1}{\rho}
\frac{\partial \rho}{\db} \ . \nonumber 
\eea
With Eqs. (22), (28) and (29), the last three equations  become 
\bea
\dtp&=&-\frac{c_n^2}{v_n a^{n-2}} \biggl( \sza + \frac{\sigma^2}{2}
\frac{\partial \szdpr}{\da}\biggr)\frac{\partial}{\da}
+\frac{\kappa_0}{v_n a^n}
\frac{\partial}{\db} \ , \nonumber \\
F_a&=&\frac{c_n^2}{v_n a^{n-2}} 
\biggl[ \hbar^2\biggl(  \frac{\partial \varphi}{\da}
+ \frac{1}{2a} \biggr) - \sigma^2 \szpr \frac{\partial \szpr}{\da}   
\biggr]\ , \nonumber \\
F_\phi&=&\frac{\sigma^2}{v_n a^n}\szpr \ .   
\eea
In order to compare Eqs. (37) , we need following explicit expressions:
\bea
\szpr&=&\frac{\epsilon_a}{2n c_n} 
{\rm ln} \left({\revz -\kappa_0 \over \revz +\kappa_0} \right) 
+ \pb \ , \nonumber \\
\sza&=&\frac{\epsilon_a}{c_n a} \revz \ , \qquad\qquad\qquad 
\frac{\partial \szpr}{\da} = \frac{\epsilon_a \kappa_0}{c_n a \revz} \ ,
\nonumber \\
\frac{\partial \szdpr}{\da}
&=&\frac{\epsilon_a \ev a^{2n-1}}{c_n (\ev a^{2n} + \kappa_0^2 )^{\frac{3}{2}}}
\ , \qquad\qquad\qquad \frac{\partial \varphi}{\da} = 
- \frac{n\ev a^{2n-1}}{2(\ev a^{2n} + \kappa_0^2)}  \ . 
\eea

The WKB approximation for the background wave function is estimated in the 
appendix, and it requires (A2), which means  
\be
\hbar \Biggl| \sza \Biggr| 
\gg \hbar^2 \Biggl| \frac{\partial \varphi}{\da} \Biggr| \ .
\ee
When $a$ is large enough to satisfy (A4), the last equation in (38) gives 
$\frac{\partial \varphi}{\partial a} \simeq -\frac{n}{2 a} \ .$  
Hence we can obtain 
\be
\hbar \Biggl| \sza \Biggr| 
\gg \hbar^2 \Biggl| \frac{\partial \varphi}{\da} +\frac{1}{2a} \Biggr| \ , 
\ee
and we can neglect the first term in $F_a$ compared to the first term in 
$\dtp$ .

Next let us examine the condition that the $\sigma^2$-term in $F_a$ 
can be neglected than the first term in $\dtp$ . 
We must require 
\be
\hbar \Biggl| \sza \Biggr| 
\gg  \sigma^2 \Biggl| \szpr \frac{\partial \szpr}{\da}  \Biggr|  \ , 
\ee
and this means when $a$ is large as in (A4) 
\be
\hbar \ev a^{2n}  \gg  \sigma^2 | \kappa_0 | \cdot 
\Biggl| - \frac{\epsilon_a \kappa_0}{n c_n \sqrt{\ev} a^n} +\pb \Biggr| \ , 
\ee
where we have used Eqs. (38). 
If $|\pb|$ is small as in (A9) , this condition (42) yields 
\be
a \gg \Biggl( \frac{\sigma^2 \kappa_0^2}{\hbar c_n \ev^{\frac{3}{2}}} 
 \Biggr)^\frac{1}{3n}
\sim \left( \frac{
\bigl( \frac{\sigma |\kappa_0 |}{\hbar} \bigr)^{\frac{2}{3}} 
\lp^{\frac{n-1}{3}} \lh}{v_n}  
\right)^\frac{1}{n} \ . 
\ee
Here we have used the definition of $\lp$ and $\lh$ in the appendix. 
Unless $|\pb|$ is small as in (A10), we need
\be
a \gg  \left( \frac{\sigma \sqrt{\frac{|\kappa_0| |\pb|}{\hbar}}}{\sqrt{\ev}} 
\right)^{\frac{1}{n}} \sim 
\left( \frac{\frac{\sigma}{\sqrt{\hbar}} 
\sqrt{\frac{|\kappa_0| |\pb|}{\hbar}}\lp^{\frac{n-1}{2}} \lh}{v_n} 
\right)^{\frac{1}{n}}\ . 
\ee

Finally consider the condition 
\be
\hbar |\kappa_0| \gg \sigma^2 |\szpr| \ , 
\ee
so that we can neglect $F_\phi$ compared to the last term in $\dtp$ . 
Suppose $a$ is large as in (A4), 
the relation (45) means 
\be
\hbar| \kappa_0 | \gg  \sigma^2  
\Biggl| - \frac{\epsilon_a \kappa_0}{n c_n \sqrt{\ev} a^n} +\pb \Biggr| \ . 
\ee
When $|\pb|$ is small as in (A9), this requires 
$\hbar |\kappa_0 | \gg \frac{\sigma^2 |\kappa_0 |}{c_n \sqrt{\ev} a^n}$ , 
which is satisfied in the region of (A7).
When $|\pb|$ is not small as in (A10), we need 
\be
\hbar |\kappa_0| \gg \sigma^2 |\pb| \ .  
\ee
This is consistent with (A10) in  the region of (A7). 

If $a$ is large enough to satisfy (A4) - (A7), (A11), (A14), (43) and (44)
and if $|\pb|$ is small enough 
to satisfy (47), $F_a , F_\phi$ can be neglected than $\dtp$ . 
In the case when the WKB approximation for the total wave function with 
$\psi$ is well justified, the last terms in the first equation of (36) 
can be also neglected, and we obtain an approximate Schr\"odinger 
equation for $\pq$ , namely 
\be
i \hbar\frac{\partial \psi}{\partial \tp} \simeq \HQ \psi \ . 
\ee

\section{Summary}
We considered a wave packet in quantum cosmology that is a superposition of the 
WKB wave functions, each of which has a definite momentum of a 
 background scalar field $\pb$ . 
We showed that it is seriously difficult to trace the formalism of 
the WKB time and to define a time variable for the Schr\"odinger equation 
of a quantum matter field 
$\pq$ without introducing a complex value for a time. 
Then we introduced a background phase time $\tp$ which is real 
and calculated its explicit 
expression when $a$ is large and $\sigma$ is small.    
It has been shown that $\tp$ is a smooth extension of $\tw$ 
which is derived by a WKB form wave function  
and they become identical in the narrow limit of the wave packet. 
We found that, when a quantum matter field $\pq$ is coupled to the 
system, an approximate 
Schr\"odinger equation for $\pq$ holds with respect to $\tp$ in a region 
where the size $a$ of the universe  is large and $|\pb|$ is small.

\newpage
\section*{Appendix : Estimation of Approximations} 
We start from the action 
$$
S = \int d^{n+1} x \sqrt{-g} \biggl[ \frac{R-2\Lambda}{16\pi G}
- \frac{g^{\mu\nu}}{2} \partial_\mu \pb \partial_\nu \pb 
+{\cal L}_{\scriptscriptstyle Q} \biggr]  \ , 
\eqno{(A1)}
$$
where $\Lambda$ is a cosmological constant and 
${\cal L}_{\scriptscriptstyle Q}$ is a Lagrangian 
density for $\pq$ . This action yields the Wheeler-DeWitt equation (1). 
>From Eq. (A1) we can see that the dimension of the scalar field is 
$\pb \sim \sqrt{\hbar} l^{\frac{1-n}{2}}$ , 
where $l$ has a dimension of length. 
Since $\kappa$ is the momentum for $\pb$ , 
$\kappa \sim \sqrt{\hbar} l^{\frac{n-1}{2}}$ , 
and $\kappa$ and  $\sigma$ have the same dimension. 
In our model the Planck length and the Hubble length are defined as 
$\lp = ( G \hbar )^{\frac{1}{n-1}}$ and 
$\lh = \frac{1}{\sqrt{\hat \Lambda}}$ , 
${\hat \Lambda} = \frac{2 \Lambda}{n(n-1)} $ , respectively. 
Then we have 
$\hbar c_n \sim \sqrt{\hbar} \lp^{\frac{n-1}{2}}$ and 
$\sqrt{\ev} \sim \sqrt{\hbar} v_n \lh^{-1}  \lp^{\frac{1-n}{2}}$ . 

Now let us begin to estimate approximations used in \S 2 and \S 3. 
In order for the WKB approximation to be well justified, 
$\Phi_0$ should vary slower than $\sz$ , which means 
$$
\frac{1}{\hbar} \Biggl| \sza \Biggr| 
\gg  \Biggl| \frac{1}{\pz} \frac{\partial \pz}{\da} \Biggr| \ , \qquad\qquad 
\frac{1}{\hbar} \Biggl| \szb \Biggr| 
\gg  \Biggl| \frac{1}{\pz} \frac{\partial \pz}{\db} \Biggr| \ . 
\eqno{(A2)}
$$
Since $\phzb = 0$ when $\gamma = 0$ , the second condition in (A2) is 
satisfied automatically. 
Using Eqs. (22) and (38), we find that the first condition of (A2) is 
$$
(\ev a^{2n} + \kappa_0^2 )^{\frac{3}{2}} \gg \hbar c_n \ev a^{2n} \ . 
\eqno{(A3)}
$$
To make discussion easy let us consider the case when $a$ is large 
enough to satisfy 
$$
a \gg \Biggl( \frac{|\kappa_0 |}{\sqrt{\ev}}  \Biggr)^\frac{1}{n}
\sim \left( \frac{ \frac{|\kappa_0 |}{\sqrt{\hbar}} 
\lp^{\frac{n-1}{2}} \lh}{v_n}  
\right)^\frac{1}{n} \ . 
\eqno{(A4)}
$$
In this case (A3) becomes 
$$
a \gg \Biggl( \frac{\hbar c_n}{\sqrt{\ev}}  \Biggr)^\frac{1}{n}
\sim \left( \frac{  \lp^{n-1} \lh}{v_n}  \right)^\frac{1}{n} \ . 
\eqno{(A5)}
$$
So this condition is necessary for the WKB approximation.

Second we consider the neglect of the higher terms than $(\kappa-\kappa_0)^2$ 
in the expansion of $f(\kappa)$ . This requires 
$|\fdpr(\kappa_0) (\kappa-\kappa_0)^2| \gg 
|f^{\prime\prime\prime}(\kappa_0)(\kappa-\kappa_0)^3|$ 
when $|\kappa-\kappa_0| \ \stackrel{<}{\sim}\ \sigma $ . This condition is satisfied when 
$$
\Biggl| \frac{\fdpr(\kappa_0)}{f^{\prime\prime\prime}(\kappa_0)} \Biggr|
\gg |\kappa-\kappa_0 | \simeq \sigma \ , 
$$
which means in the case of (A4) that 
$$
a \gg \Biggl( \frac{\sigma |\kappa_0 | }{\ev}  \Biggr)^\frac{1}{2n}
\sim \left( \frac{ \sqrt{\frac{\sigma |\kappa_0 | }{\hbar} }
\lp^{\frac{n-1}{2}} \lh}{v_n}  
\right)^\frac{1}{n} \ . 
\eqno{(A6)}
$$
Here we have used Eqs. (22) and (38).

Next let us examine the condition (24). In the case of (A4) we obtain 
from Eqs. (22) and (38) that 
$\sigma^2 |\fdpr| \simeq \frac{\sigma^2}{\hbar} |\szdpr| \simeq 
\frac{\sigma^2}{\hbar n c_n \sqrt{\ev} a^n} $ . Therefore (24) requires 
$$
a \gg \Biggl( \frac{\sigma^2}{\hbar c_n \sqrt{\ev}}  \Biggr)^\frac{1}{n}
\sim \left( \frac{\bigl( \frac{\sigma}{\sqrt{\hbar}} \bigr)^2 \lh}{v_n}  
\right)^\frac{1}{n} \ . 
\eqno{(A7)}
$$

Finally we estimate conditions (26), (27). 
In the case of (A4) the condition (26) gives 
$$
\frac{\sqrt{\ev} a^n}{c_n | \kappa_0 |} \gg    
\Biggl| - \frac{\epsilon_a \kappa_0}{n c_n \sqrt{\ev} a^n} +\pb \Biggr| \ . 
\eqno{(A8)}
$$
When $|\pb|$ is small and satisfies 
$$
|\pb| \  {\mathop{<}_\sim}\ \; \frac{| \kappa_0 |}{n c_n \sqrt{\ev} a^n} \ , 
\eqno{(A9)}
$$
(A8) is satisfied by (A4). When $|\pb|$ is not small, that is 
$$
|\pb| \  {\mathop{>}_\sim}\ \;  \frac{| \kappa_0 |}{n c_n \sqrt{\ev} a^n} \ , 
\eqno{(A10)}
$$
(A8) means 
$$
a \gg \Biggl( \frac{c_n |\kappa_0 | |\pb| }{\sqrt{\ev}}  \Biggr)^\frac{1}{n}
\sim \left( \frac{ \frac{|\kappa_0 | |\pb|}{\hbar} 
\lp^{n-1} \lh}{v_n}  
\right)^\frac{1}{n} \ . 
\eqno{(A11)}
$$
In the case of (A4) the condition (27) requires 
$$    
\Biggl| - \frac{\epsilon_a \kappa_0}{n c_n \sqrt{\ev} a^n} +\pb \Biggr| 
\gg \frac{\hbar}{\sqrt{\ev} a^n}\ . 
\eqno{(A12)}
$$
When $|\pb|$ is small as in (A9), the condition (A12) becomes 
$$
\frac{|\kappa_0|}{\sqrt{\hbar}} 
\gg \sqrt{\hbar} c_n \sim \lp^{\frac{n-1}{2}} \ , 
\eqno{(A13)}
$$
and when $|\pb|$ satisfies (A10), the condition (A12) yields  
$$
a \gg \Biggl( \frac{\hbar}{\sqrt{\ev} |\pb|}  \Biggr)^\frac{1}{n}
\sim \left( \frac{  
\lp^{\frac{n-1}{2}} \lh}{v_n \frac{|\pb |}{\sqrt{\hbar}}}  
\right)^\frac{1}{n} \ . 
\eqno{(A14)}
$$

Since $v_n a^n$ is the spatial volume of our model, (A5) may be satisfied 
in the semiclassical region. If $\sigma$ is small and 
$\frac{|\kappa_0|}{\sqrt{\hbar}}$ is bigger than $l_p^{\frac{n-1}{2}}$ 
as in (A13) but is not too big, (A4),(A6) and (A7) do not impose too 
severe restriction on $a$ . 
When $|\pb| \cdot |\kappa_0| \simeq \hbar$ , which is allowed by 
the uncertainty principle, the conditions (A11) and (A14) are equivalent 
to (A5) and (A4), respectively. 
So we can think that, when $a$ is large and $\sigma$ is small, all the 
conditions in this appendix can be satisfied consistently.

\newpage

\end{document}